\documentclass[amsmath,amssymb,showkeys,superscriptaddress,floatfix,prd,10pt]{revtex4}
\usepackage{graphicx,epstopdf}
\usepackage{subfig}
\usepackage{xcolor}
\usepackage[colorlinks=true, urlcolor=blue, linkcolor=blue, citecolor=green]{hyperref}
\def\pslash{p\!\!\!\slash }
\def\qslash{q\!\!\!\slash }
\def\xslash{x\!\!\!\slash }
\def\eslash{\varepsilon\!\!\!\slash }

\begin{document}

\title{Magnetic dipole moments of the hidden-charm pentaquark
states: $P_c(4440)$, $P_c(4457)$ and $P_{cs}(4459)$ }

\author{Ula\c{s} \"{O}zdem}%
\email[]{ulasozdem@aydin.edu.tr}
\affiliation{Health Services Vocational School of Higher Education, Istanbul Aydin University, Sefakoy-Kucukcekmece, 34295 Istanbul, Turkey}

\date{\today}
 
\begin{abstract}
In this work, we employ the light-cone QCD sum rule to calculate the magnetic dipole moments  of the  $P_c(4440)$, $P_c(4457)$ and $P_{cs}(4459)$ pentaquark states by considering them as the diquark-diquark-antiquark and molecular  pictures  with quantum numbers  $J^P = \frac{3}{2}^-$, $J^P = \frac{1}{2}^-$ and $J^P = \frac{1}{2}^-$, respectively.
%
In the analyses, we use the diquark-diquark-antiquark and molecular form of interpolating currents, and photon distribution amplitudes to obtain the magnetic dipole moment of pentaquark states.
Theoretical examinations on magnetic dipole moments of the hidden-charm pentaquark states, are essential as their results can help us better figure out their substructure and the dynamics of the QCD as the theory of the strong interaction.
As a by product, we extract the electric quadrupole and magnetic octupole moments of the $P_c(4440)$ pentaquark. These values show a non-spherical charge distribution.
\end{abstract}
\keywords{Pentaquarks, magnetic moment, $P_c(4440)$, $P_c(4457)$,  $P_{cs}(4459)$, diquark-diquark-antiquark picture, molecular picture}

\maketitle

\section{Motivation}\label{motivation}
With the experimental discovery of the exotic X(3872) state, i.e., state that cannot be interpreted by the conventional meson or baryon picture, a new era began in high energy physics. 
 Since the discovery of this state, numerous exotic states have been added to the family of particles that have been experimentally reported. %
Study of exotic particles that are composed of tetraquark, pentaquark or hybrid states is among the most attractive subjects of the hadron physics.
Experimental data collected with various collaborations in recent years and theoretical developments obtained in different theoretical models form the field of rapidly growing exotic studies~\cite{Chen:2016qju,Ali:2017jda,Esposito:2016noz,Olsen:2017bmm,Lebed:2016hpi,Guo:2017jvc,Albuquerque:2018jkn,Brambilla:2019esw,Liu:2019zoy,Agaev:2020zad, Dong:2021juy}.

In 2019, LHCb collaboration  reported three new narrow pentaquark states~\cite{Aaij:2019vzc} as
%
\begin{align}
P_c(4312): \mbox{M}&=4311.9 \pm 0.7 ^{+6.8}_{-0.6}~\mbox{MeV},~~~
\Gamma = 9.80 \pm 2.7^{+3.7}_{-4.5} ~\mbox{MeV},\nonumber\\
P_c(4440): \mbox{M}&=4440.3 \pm 1.3 ^{+4.1}_{-4.7} ~\mbox{MeV},~~~
 \Gamma = 20.6 \pm 4.9^{+8.7}_{-10.1} ~\mbox{MeV},\nonumber\\
P_c(4457): \mbox{M}&=4457.3 \pm 0.6 ^{+4.1}_{-1.7} ~\mbox{MeV},~~~
      \Gamma = 6.40 \pm 2.0^{+5.7}_{-1.9} ~\mbox{MeV}\nonumber.
\end{align}
 
 Very recently, the LHCb Collaboration reported a pentaquark state with strangeness, $P_{cs}(4459)$, 
 in the invariant mass spectrum of $J/\psi\Lambda$ in the $\Xi_b^0 \rightarrow J/\psi\,\Lambda\,K^-$ decay~\cite{Aaij:2020gdg}:
 \begin{align}
P_{cs}(4459): \mbox{M}&=4458.8 \pm 2.7 ^{+4.7}_{-1.1}~\mbox{MeV},~~~
\Gamma = 17.3 \pm 6.5^{+8.0}_{-5.7}~ \mbox{MeV}.
\end{align}
 But, spin and parity of the $P_{cs}(4459)$ state have not been determined yet.
 As regards to their decay products, one can easily conclude that these newly discovered four states consist of at least five  quarks, $\bar{c}cuud$ or $\bar{c}cuds$, therefore they are perfect candidates of hidden-charm pentaquark states.
 After the experimental discovery
several phenomenological models have been adopted to calculate the spectroscopic parameters, decays and production mechanisms of the pentaquarks, like the QCD sum rule, the meson-exchange model, the quark delocalization model, and so on
~\cite{Chen:2019asm,Chen:2019bip,Liu:2019tjn,He:2019ify,Xiao:2019mvs,Guo:2019kdc,Xiao:2019aya,Zhang:2019xtu,Wu:2019rog,Wang:2019ato,Xu:2020gjl,Peng:2020gwk,Shimizu:2019ptd,Zhu:2019iwm,Wang:2019got,Wang:2018waa,Cheng:2019obk,Yamaguchi:2019seo,Pan:2019skd,Liu:2019zvb,Fernandez-Ramirez:2019koa,Eides:2019tgv,Guo:2019fdo,Cao:2019kst,Wang:2019dsi,Wang:2019krd,Xiao:2020frg,Mutuk:2019snd,Yang:2020twg,Dong:2020nwk,Wang:2020rdh,Du:2019pij,Wang:2019spc,Azizi:2020ogm,Chen:2020opr,Chen:2020uif, Liu:2020hcv,Peng:2020hql,Chen:2020kco,Wang:2020eep,Azizi:2021utt,Zhu:2021lhd}.
However, the substructure of these states are not determined yet. In other words, in order to understand the internal structure and nature of these particles, different properties should be studied besides their decay channels and spectroscopic properties.
For instance, investigating their electromagnetic form factors may provide important insights on this point.

Electromagnetic form factors or multi-pole moments of hadrons are important parameters in study of their electromagnetic structure, and also they can ensure important knowledge about the dynamics of the QCD at low energy region. 
 Electromagnetic multi-pole moments, especially magnetic dipole moment, are also a crucial part in the calculation of J$/\psi$ photo-production cross sections, which can provide an independent analysis of the hidden-charm pentaquark states.
 Investigating electromagnetic features of exotic resonances is relatively new topic. However, in the literature there are a few studies where the electromagnetic properties of the hidden-charm pentaquark states are investigated~\cite{Wang:2016dzu, Ozdem:2018qeh, Ortiz-Pacheco:2018ccl,Ozdem:2021btf,Xu:2020flp}.
 In Ref.~\cite{Wang:2016dzu}, the magnetic dipole moment of the hidden-charm pentaquark states have been extracted in the diquark-diquark-antiquark,  diquark-triquark  and molecular configuration with $J^P = \frac{1}{2}^{\pm}, \frac{3}{2}^{\pm}, \frac{5}{2}^{\pm} \mbox{and}~ \frac{7}{2}^{+}$ quantum numbers in the different color-flavor structure.  
 In Ref.~\cite{Ozdem:2018qeh}, the  magnetic dipole,  electric quadrupole and magnetic octupole moments  of the $P_c(4380)$ pentaquark with  $J^P = \frac{3}{2}^{-}$  quantum numbers have been obtained in the diquark-diquark-antiquark and molecular pictures in the framework of the  light-cone QCD sum rule (LCSR). 
 In Ref.~\cite{ Ortiz-Pacheco:2018ccl}, they  acquired the the ground state of hidden-charm pentaquarks with $J^P = \frac{3}{2}^{-}$  quantum numbers and their associated magnetic dipole moments and electromagnetic couplings, of interest to pentaquark photoproduction experiments in the framework of the  constituent quark model. 
 In Ref.~\cite{Ozdem:2021btf}, the magnetic dipole moment of the $P_c(4312)$ pentaquark state have been extracted  in the molecular and diquark-diquark-antiquark pictures via the LCSR with  $J^P = \frac{1}{2}^{-}$  quantum numbers. 
 In Ref. \cite{Xu:2020flp}, they achieved the magnetic dipole moment of the $P_c(4312)$ pentaquark state in the  $\Sigma_c \bar D$ molecular picture by means of the QCD sum rule (QCDSR) in the external weak electromagnetic field. 
 %
 
  %
 In this study, the magnetic dipole  moments of the pentaquark states $P_c(4440)$, $P_c(4457)$  and  $P_{cs}(4459)$ (hereafter we will show these states as $P_{c1}$, $P_{c2}$ and  $P_{cs}$, respectively) are obtained by employing the diquark-diquark-antiquark and molecular  interpolating currents within the LCSR. 
 The LCSR is a powerful tool to obtain qualitative and quantitative information about hadron properties. In this method, hadrons are analytically studied by matching their behaviors at high and low energies. This matching provides characteristic properties of hadrons from QCD. The information obtained by LCSR has been used as an input to other theoretical and experimental approaches for a long time~\cite{Chernyak:1990ag, Braun:1988qv, Balitsky:1989ry}. 
  %
  %

The rest of paper is structured as follows. In Sect. \ref{formalism}, the method used for the calculations are described. In Sect. \ref{numerical}, we carry out numerical analysis of the acquired LCSR for the magnetic dipole moment of pentaquark states. 
The explicit expressions of the magnetic moment of the $P_{c1}$ pentaquark in the diquark-diquark-antiquark picture is presented in Appendix A.

\section{Magnetic dipole  moments of the pentaquark states via LCSR}\label{formalism}
In this section, we briefly introduce the LCSR method used to  extract the magnetic dipole moments of the $P_{c1}$, $P_{c2}$ and $P_{cs}$  pentaquark states.
In order to obtain the magnetic dipole moment of the corresponding states with the QCD sum rule method, we begin by writing the correlation function suitable for the calculations.
This correlation function is obtained in two representations which are called hadronic and QCD
representations. The LCSR for the physical quantities are obtained from the matches of the coefficients of the same Lorentz structures achieved on both representations with the help of the quark-duality ansatz.

\subsection{Formalism of the \texorpdfstring{$P_{c2}$}{} and \texorpdfstring{$P_{cs}$}{} states}

 The current correlation function for the $P_{c2}$ and $P_{cs}$ pentaquark states is written as
\begin{equation}
 \label{edmn01}
\Pi(p,q)=i\int d^{4}xe^{ip\cdot x}\langle 0|\mathcal{T}\{J_{P_{c2(cs)}}(x)
\bar J_{P_{c2(cs)}}(0)\}|0\rangle_{\gamma}, 
\end{equation}%
where $J_{P_{c2(cs)}}(x)$ is the interpolating current of $P_{c2}$ or $P_{cs}$ pentaquark state. In the diquark-diquark-antiquark and molecular pictures with quantum number $J^P = \frac{1}{2}^-$, they are written as
\begin{eqnarray} \label{intcur}
 J_{P_{c2}}^{Di}(x)&=&\varepsilon^{abc}\varepsilon^{ade}\varepsilon^{bfg}\big[  u^T_d(x) C\gamma_{5} d_e(x)\,u^T_f(x) 
  C\gamma_{\mu} c_g(x)\, \gamma_5\gamma^{\mu}C\bar{c}^{T}_{c}(x)\big],
 \nonumber\\
 J_{P_{c2}}^{Mol}(x)&=&\big[\bar c_d(x)\gamma_\mu u_d(x)\big] \gamma^{\mu}\gamma_5\big[\varepsilon^{abc}  u^T_a(x) C \gamma_{\nu} d_b(x) \gamma^{\nu} \gamma_5 c_{c}(x)\big],
\\
 \nonumber\\
 J_{P_{cs}}^{Di}(x)&=&\varepsilon^{abc}\varepsilon^{ade}\varepsilon^{bfg}\big[  u^T_d(x) C\gamma_5 d_e(x)\,s^T_f(x) 
 C\gamma_5 c_g(x)\, C\bar{c}^{T}_{c}(x)\big],\nonumber\\
  J_{P_{cs}}^{Mol}(x)&=&\big[\bar c_d(x)\gamma_\mu u_d(x)\big] \gamma^{\mu}\gamma_5\big[\varepsilon^{abc}  d^T_a(x) C \gamma_{5} s_b(x) c_{c}(x)\big],
 \end{eqnarray}
where $C$ is the charge conjugation matrix; and $a$, $b$... are color indices.

As we mentioned at the beginning of this section, in LCSR studies we need to evaluate the correlation function at both hadron and quark-gluon degrees of freedom. At the hadron level, we embedding a complete set of intermediate pentaquark states into the correlation function to acquire the hadronic representation, and isolate the ground state pentaquark states, and get the results:
 \begin{align}\label{edmn02}
\Pi^{Had}(p,q)&=\frac{\langle0\mid J_{P_{c2(cs)}} \mid
{P_{c2(cs)}}(p, s) \rangle}{[p^{2}-m_{P_{c2(cs)}}^{2}]}
\langle {P_{c2(cs)}}(p, s)\mid
{P_{c2(cs)}}(p+q, s)\rangle_\gamma 
\frac{\langle {P_{c2(cs)}}(p+q, s)\mid
\bar{J}_{P_{c2(cs)}} \mid 0\rangle}{[(p+q)^{2}-m_{{P_{c2(cs)}}}^{2}]}+...,
\end{align}

The matrix element  $\langle
{P_{c2(cs)}}(p, s)\mid {P_{c2(cs)}}(p+q, s)\rangle_\gamma$ entering Eq. (\ref{edmn02}) can be parameterized in terms of Lorentz invariant form factors as follows:
%
\begin{align}
\langle {P_{c2(cs)}}(p, s)\mid {P_{c2(cs)}}(p+q, s)\rangle_\gamma &=\varepsilon^\mu\,\bar u(p, s)\Big[\big[f_1(q^2)
+f_2(q^2)\big] \gamma_\mu +f_2(q^2)
\frac{(2p+q)_\mu}{2 m_{P_{c2(cs)}}}\Big]\,u(p+q, s) \label{edmn04},
\end{align}
%
where 
$\varepsilon$ and q  are the polarization vector and momentum of the photon, respectively. 

Substituting Eq.~(\ref{edmn04}) in Eq. (\ref{edmn02}) for hadronic side we get
\begin{align}
\label{edmn05}
\Pi^{Had}(p,q)=&\lambda^2_{P_{c2(cs)}}\gamma_5 \frac{\Big(\pslash+m_{P_{c2(cs)}} \Big)}{[p^{2}-m_{{P_{c2(cs)}}}^{2}]}\varepsilon^\mu \Bigg[\big[f_1(q^2) %
+f_2(q^2)\big] \gamma_\mu
+f_2(q^2)\, \frac{(2p+q)_\mu}{2 m_{P_{c2(cs)}}}\Bigg]  \gamma_5 
\frac{\Big(\pslash+\qslash+m_{P_{c2(cs)}}\Big)}{[(p+q)^{2}-m_{{P_{c2(cs)}}}^{2}]}, 
\end{align}
with $\langle0\mid J_{P_{c2(cs)}}\mid P_{c2(cs)}(p, s)\rangle = \lambda_{P_{c2(cs)}} \gamma_5 \, u(p,s)$.
The Lorentz invariant form factors $f_1(q^2)$ and $f_2(q^2)$ are related to the magnetic form factor by the relation,
\begin{align}
G_M(q^2) = f_1(q^2)+f_2(q^2).
\end{align}
At $q^2 =0$,  the magnetic form factor is acquired associated with the functions  $f_1(q^2)$ and $f_2(q^2)$ as:
\begin{align}
G_M(0) = f_1(0)+f_2(0).
\end{align}
The magnetic dipole  moment, $\mu_{P_{c2(cs)}}$,  is defined in the following way:
\begin{align}
 \mu_{P_{c2(cs)}} =  G_M(0) \, \frac{e}{2\,m_{P_{c2(cs)}}}.
\end{align}
We observe from Eq.(\ref{edmn05}) that the correlation function contains many structures, any of them can be selected in obtaining magnetic dipole moment of the $P_{c2(cs)}$  pentaquark, and thus we decide on the structure $\eslash\qslash$. As a result, the correlation function can be expressed with regards to the magnetic dipole moment of  $P_{c2(cs)}$   pentaquark as,
\begin{align}
\label{edmn07}
\Pi^{Had}(p,q)&= \frac{\lambda^2_{P_{c2(cs)}}\,m_{P_{c2(cs)}}}{[(p+q)^2-m^2_{P_{c2(cs)}}]} \,\mu_{P_{c2(cs)}}\,
\frac{1}{[p^2-m^2_{P_{c2(cs)}}]}.
\end{align}
At the QCD level, we contract the relevant quark fields in the correlation function with the help of the Wick's theorem, 
\begin{widetext}
\begin{align}
\label{edmn11}
\Pi^{QCD-Di}_{P_{c2}}(p,q)&=i\,\varepsilon^{abc}\varepsilon^{a^{\prime}b^{\prime}c^{\prime}}\varepsilon^{ade}
\varepsilon^{a^{\prime}d^{\prime}e^{\prime}}\varepsilon^{bfg}
\varepsilon^{b^{\prime}f^{\prime}g^{\prime}} \int d^4x e^{ip\cdot x}\langle 0|
\bigg\{ \nonumber\\
&   Tr\Big[\gamma_5 S_d^{ee^\prime}(x) \gamma_5 \tilde S_u^{dd^\prime}(x)\Big]
Tr\Big[\gamma_\mu S_c^{gg^\prime}(x) \gamma_\nu \tilde S_u^{ff^\prime}(x)\Big]  
 \nonumber\\
& -  Tr \Big[\gamma_5 S_d^{ee^\prime}(x) \gamma_5 \tilde S_u^{fd^\prime}(x) 
\gamma\mu S_c^{gg^\prime}(x) \gamma_\nu 
\tilde S_u^{df^\prime}(x)\Big]  \bigg \}  
\Big(\gamma_5 \gamma^{\mu} S_c^{c^{\prime}c}(-x) \gamma ^{\nu} \gamma_5\Big)
|0 \rangle_\gamma ,\\
%
\Pi^{QCD-Mol}_{P_{c2}}(p,q)&=i\,\epsilon^{abc}\epsilon^{a^{\prime}b^{\prime}c^{\prime}}\,\int d^{4}xe^{ip\cdot x}
\langle 0|
\Bigg\{\mathrm{Tr}\Big[\gamma_{\mu} S_{u}^{dd^{\prime}}(x) 
\gamma_{\alpha}S_c^{d^{\prime}d}(-x)\Big] 
\mathrm{Tr}\Big[\gamma_{\nu}S_d^{bb^{\prime}}(x)\gamma_{\beta}\widetilde{S}_u^{aa^{\prime}}(x)]
\nonumber \\ 
&-
\mathrm{Tr}\Big[\gamma_{\mu} S_{u}^{da^{\prime}}(x) 
\gamma_{\beta}\widetilde{S}_d^{bb^{\prime}}(x)\Big]
\mathrm{Tr}\Big[\gamma_{\nu}{S}_{u}^{ad^{\prime}}(x)\gamma_{\alpha}S_c^{d^{\prime}d}(-x)\Big]
\Bigg\} \Big( \gamma^{\mu}\gamma^{\nu}S_c^{cc^{\prime}}(x)\gamma^{\alpha}\gamma^{\beta}\Big)\Bigg
|0 \rangle_\gamma,
 \label{abcd}
\end{align}
\end{widetext}

\begin{widetext}
\begin{align}
\label{edm12}
\Pi^{QCD-Di}_{P_{cs}}(p,q)=&-i\,\varepsilon^{abc}\varepsilon^{a^{\prime}b^{\prime}c^{\prime}}\varepsilon^{ade}
\varepsilon^{a^{\prime}d^{\prime}e^{\prime}}\varepsilon^{bfg}
\varepsilon^{b^{\prime}f^{\prime}g^{\prime}}  \int d^4x 
 e^{ip\cdot x} 
 \nonumber\\
&       \langle 0|
 Tr\Big[\gamma_5 S_d^{ee^\prime}(x) \gamma_5 \tilde S_u^{dd^\prime}(x)\Big] 
Tr\Big[\gamma_5 S_c^{gg^\prime}(x) \gamma_5 \tilde S_s^{ff^\prime}(x)\Big]
\tilde S_c^{c^{\prime}c}(-x)
|0 \rangle_\gamma ,\\
%
\Pi_{P_{cs}}^{\mathrm{QCD-Mol}}(p,q)=&-i\,\epsilon^{abc}\epsilon^{a^{\prime}b^{\prime}c^{\prime}}\,\int d^{4}xe^{ip\cdot x}\nonumber\\
&\langle 0|
\mathrm{Tr}\Big[\gamma_{\nu} S_{u}^{dd^{\prime}}(x) 
\gamma_{\mu}S_c^{d^{\prime}d}(-x)\Big]
\mathrm{Tr}\Big[\gamma^{\nu}\widetilde{S}_s^{bb^{\prime}}(x)\gamma^{\mu}S_d^{aa^{\prime}}(x) \Big]
S_c^{cc^{\prime}}(x)
|0 \rangle_\gamma,
 \label{eq:CorrF2Pc1}
\end{align}
\end{widetext}
where
\begin{equation*}
\widetilde{S}_{c(q)}^{ij}(x)=CS_{c(q)}^{ij\mathrm{T}}(x)C,
\end{equation*}%
with $S_{q(c)}(x)$ being the full light and charm quark propagators.
The relevant propagators are given as~\cite{Balitsky:1987bk}
\begin{widetext}
\begin{align}
\label{edmn13}
S_{q}(x)&=i \frac{{\xslash}}{2\pi ^{2}x^{4}} 
- \frac{ \bar qq }{12} \Big(1-i\frac{m_{q} \xslash}{4}   \Big)
- \frac{ \bar qq }{192}m_0^2 x^2  \Big(1
-i\frac{m_{q} \xslash}{6}   \Big)
-\frac {i g_s }{32 \pi^2 x^2} ~G^{\mu \nu} (x) \Big[\rlap/{x} 
\sigma_{\mu \nu} +  \sigma_{\mu \nu} \rlap/{x}
 \Big],\\
\nonumber\\
\label{edmn14}
S_{c}(x)&=\frac{m_{c}^{2}}{4 \pi^{2}} \Bigg[ \frac{K_{1}\Big(m_{c}\sqrt{-x^{2}}\Big) }{\sqrt{-x^{2}}}
+i\frac{{\xslash}~K_{2}\Big( m_{c}\sqrt{-x^{2}}\Big)}
{(\sqrt{-x^{2}})^{2}}\Bigg]
-\frac{g_{s}m_{c}}{16\pi ^{2}} \int_0^1 dv\, G^{\mu \nu }(vx)\Bigg[ (\sigma _{\mu \nu }{\xslash}
  +{\xslash}\sigma _{\mu \nu })
  \nonumber\\
  &\times \frac{K_{1}\Big( m_{c}\sqrt{-x^{2}}\Big) }{\sqrt{-x^{2}}}
+2\sigma_{\mu \nu }K_{0}\Big( m_{c}\sqrt{-x^{2}}\Big)\Bigg],
\end{align}%
\end{widetext}
where $K_i$ are modified the second kind Bessel functions and $G^{\mu\nu}$ is the gluon field strength tensor.

The QCD representation of the correlation function can be obtained with the help of photon distribution amplitudes (DAs) according to quark-gluon properties and after performing the Fourier transform to transfer the calculations to the momentum space.

As a final step, by applying the double Borel transform on the variables $ -p^2 $ and $ (p + q)^2 $ and choosing the coefficients of the same Lorentz structures in both QCD and hadronic representations and matching them employing the quark-hadron duality approach, we obtain the desired LCSR for magnetic dipole  moments of $P_{c2}$ and $P_{cs}$ states:
  \begin{align}
\label{sonucPc2}
 &\mu_{P_{c2}}^{Di} \,\lambda^{2-Di}_{P_{c2}}\, m_{P_{c2}} = e^{\frac{m^2_{P_{c2}}}{M^2}}\, \Delta_1^{QCD},\\
\label{sonucPc2-1}
& \mu_{P_{c2}}^{Mol} \,\lambda^{2-Mol}_{P_{c2}}\, m_{P_{c2}} = e^{\frac{m^2_{P_{c2}}}{M^2}}\, \Delta_2^{QCD},\\
 & \mu_{P_{cs}}^{Di} \,\lambda^{2-Di}_{P_{cs}}\, m_{P_{cs}} = e^{\frac{m^2_{P_{cs}}}{M^2}}\, \Delta_3^{QCD}.
 \label{sonucPcs}\\
&  \mu_{P_{cs}}^{Di} \,\lambda^{2-Mol}_{P_{cs}}\, m_{P_{cs}} = e^{\frac{m^2_{P_{cs}}}{M^2}}\, \Delta_4^{QCD}.
 \label{sonucPcs-1}
 \end{align}

The $\Delta_1^{QCD}$, $\Delta_2^{QCD}$, $\Delta_3^{QCD}$ and $\Delta_4^{QCD}$ functions are quite lengthy, therefore the explicit expression of this function are not presented here.

%

\subsection{Formalism of the \texorpdfstring{$P_{c1}$}{} state}

 In this subsection we derive the LCSR for the magnetic dipole moment of the 
 $P_{c1}$ pentaquark state. For this purpose, we consider following correlation function,
\begin{equation}
 \label{Pc101}
\Pi _{\mu \nu }(p,q)=i\int d^{4}xe^{ip\cdot x}\langle 0|\mathcal{T}\{J_{\mu}^{P_{c1}}(x)
\bar J_{\nu }^{P_{c1}}(0)\}|0\rangle_{\gamma}, 
\end{equation}%
where $J_{\mu(\nu)}^{P_{c1}}$ is the interpolating current of $P_{c1}$ pentaquark with  $J^P = \frac{3}{2}^{-}$  quantum numbers. In the diquark-diquark-antiquark and molecular pictures, 
it is given as
\begin{eqnarray}
 J_{\mu}^{P_{c1}-Di}(x)&=&\varepsilon^{abc}\varepsilon^{ade}\varepsilon^{bfg}\big[  u^T_d(x) C\gamma_5 d_e(x)\,u^T_f(x) 
 C\gamma_\mu c_g(x)\, C\bar{c}^{T}_{c}(x)\big],\nonumber\\
 \label{eq:Pc102}
 J_{\mu}^{P_{c1}-Mol}(x)&=&\big[\bar c_d(x)\gamma_\mu u_d(x)\big] [\varepsilon^{abc}  u^T_a(x) C \gamma_{\nu} d_b(x) \gamma^{\nu} \gamma_5 c_{c}(x)\big]
\end{eqnarray}

The hadronic side of the correlation function is written as,
\begin{eqnarray}\label{Pc103}
\Pi^{Had}_{\mu\nu}(p,q)&=&\frac{\langle0\mid  J_{\mu}^{P_{c1}}\mid
{P_{c1}}(p)\rangle}{[p^{2}-m_{{P_{c1}}}^{2}]}\langle {P_{c1}}(p)\mid
{P_{c1}}(p+q)\rangle_\gamma 
\frac{\langle {P_{c1}}(p+q)\mid
\bar{J}_{\nu}^{P_{c1}}\mid 0\rangle}{[(p+q)^{2}-m_{{P_{c1}}}^{2}]}+...
\end{eqnarray}
The matrix element of the interpolating current 
between the vacuum and the $P_{c1}$ pentaquark is defined as
\begin{equation}\label{lambdabey}
\langle0\mid J_{\mu}^{P_{c1}}(0)\mid {P_{c1}}(p,s)\rangle=\lambda_{{P_{c1}}}u_{\mu}(p,s),
\end{equation}
where $\lambda_{{P_{c1}}}$ is the  residue $P_{c1}$ pentaquark and $u_{\mu}(p,s)$ is the Rarita-Schwinger spinor. 

The transition matrix element $\langle
{P_{c1}}(p)\mid {P_{c1}}(p+q)\rangle_\gamma$ entering Eq.
(\ref{Pc103}) can be parameterized in terms of four Lorentz invariant form factors as follows
\cite{Weber:1978dh,Nozawa:1990gt,Pascalutsa:2006up,Ramalho:2009vc}:
\begin{align}\label{matelpar}
\langle {P_{c1}}(p)\mid {P_{c1}}(p+q)\rangle_\gamma &=-e\bar
u_{\mu}(p)\Bigg[F_{1}(q^2)g_{\mu\nu}\eslash-
\frac{1}{2m_{{P_{c1}}}} 
\Big[F_{2}(q^2)g_{\mu\nu}+F_{4}(q^2)\frac{q_{\mu}q_{\nu}}{(2m_{{P_{c1}}})^2}\Big]\eslash\qslash
\nonumber\\&+
F_{3}(q^2)\frac{1}{(2m_{{P_{c1}}})^2}q_{\mu}q_{\nu}\eslash \Bigg] u_{\nu}(p+q).
\end{align}

In principle, we can obtain the final expression of the hadronic side of the correlation function using the above equations, but we encounter two difficulties: not all Lorentz structures are independent and the correlation function also includes contributions of spin- {1/2} and these undesirable contributions must be eliminated.
To remove undesirable contributions coming from the spin-1/2 particles and obtain only independent structures in the
correlation function, we apply the  ordering for Dirac
matrices as $\gamma_{\mu}\pslash\eslash\qslash\gamma_{\nu}$ and remove terms 
with $\gamma_\mu$ at the beginning, $\gamma_\nu$ at the end and those proportional to $p_\mu$ and 
$p_\nu$~\cite{Belyaev:1982cd}. As a result, using Eqs. (\ref{Pc101})-(\ref{matelpar})
the hadronic side take the form,
\begin{align}\label{final phenpart}
\Pi^{Had}_{\mu\nu}(p,q)=&\frac{\lambda_{_{{P_{c1}}}}^{2}}{[(p+q)^{2}-m_{_{{P_{c1}}}}^{2}][p^{2}-m_{_{{P_{c1}}}}^{2}]} 
\Bigg[  g_{\mu\nu}\pslash\eslash\qslash \,F_{1}(q^2) 
-m_{{P_{c1}}}g_{\mu\nu}\eslash\qslash\,F_{2}(q^2)
-
\frac{F_{3}(q^2)}{4m_{{P_{c1}}}}q_{\mu}q_{\nu}\eslash\qslash\,
-
\frac{F_{4}(q^2)}{4m_{{P_{c1}}}^3}(\varepsilon.p)q_{\mu}q_{\nu}\pslash\qslash 
\nonumber\\
&
+
\mbox{other independent structures} \Bigg].
\end{align}
The final form of the hadronic side in terms of the selected structures in momentum space is:
\begin{eqnarray}
\Pi^{Had}_{\mu\nu}(p,q)&=&\Pi_1^{Had}g_{\mu\nu}\pslash\eslash\qslash \,
+\Pi_2^{Had}g_{\mu\nu}\eslash\qslash\,+
...,
\end{eqnarray}
where $ \Pi_1^{Had} $ and $ \Pi_2^{Had} $ are functions of the form factors $ F_1(q^2) $ and  $ F_2(q^2) $, respectively; and other independent structures is represented by dots.

The magnetic, $G_{M}(q^2)$, form factor is
defined in terms of the form factors $F_{i}(q^2)$ in the following
way
 \cite{Weber:1978dh,Nozawa:1990gt,Pascalutsa:2006up,Ramalho:2009vc}:
\begin{align}
G_{M}(q^2) &= [ F_1(q^2) + F_2(q^2)] ( 1+ \frac{4}{5}
\tau ) -\frac{2}{5} [ F_3(q^2)  
+ 
F_4(q^2)] \tau ( 1 + \tau ), 
\end{align}
  where $\tau
= -\frac{q^2}{4m^2_{{P_{c1}}}}$. At $q^2=0$, the magnetic dipole moment
is obtained in terms of the functions $F_1(0)$ and $F_2(0)$ as:
\begin{eqnarray}\label{mqo1}
G_{M}(0)&=&F_{1}(0)+F_{2}(0).
\end{eqnarray}
The  magnetic dipole moment, ($\mu_{{P_{c1}}}$), is defined in the following way:
 \begin{eqnarray}\label{mqo2}
\mu_{{P_{c1}}}&=&\frac{e}{2m_{{P_{c1}}}}G_{M}(0).
\end{eqnarray}

The next step is to calculate the correlation function in  Eq.~(\ref{Pc101}) in terms of quark-gluon
parameters. When we apply the same procedures as in the previous subsection, we get the following result:
\begin{widetext}
\begin{eqnarray}
\label{QCDPc11}
\Pi^{QCD-Di}_{\mu\nu}(p,q)&=&i\,\varepsilon^{abc}\varepsilon^{a^{\prime}b^{\prime}c^{\prime}}\varepsilon^{ade}
\varepsilon^{a^{\prime}d^{\prime}e^{\prime}}\varepsilon^{bfg}
\varepsilon^{b^{\prime}f^{\prime}g^{\prime}} \int d^4x 
e^{ip\cdot x} \langle 0|
\Bigg\{
\nonumber\\
&&
  Tr\Big[\gamma_5 S_d^{ee^\prime}(x) \gamma_5 \tilde S_u^{dd^\prime}(x)\Big] 
Tr\Big[\gamma_\mu S_c^{gg^\prime}(x) \gamma_\nu \tilde S_u^{ff^\prime}(x)\Big]\tilde S_c^{c^{\prime}c}(-x) \nonumber\\
&& -  Tr \Big[\gamma_5 S_d^{ee^\prime}(x) \gamma_5 \tilde S_u^{fd^\prime}(x) 
\gamma_\mu S_c^{gg^\prime}(x) \gamma_\nu 
\tilde S_u^{df^\prime}(x)\Big] \tilde S_c^{c^{\prime}c}(-x)\Bigg \}
|0 \rangle_\gamma,
\\
%
\Pi_{\mu \nu }^{\mathrm{QCD-Mol}}(p,q)&=&-i\,\epsilon^{abc}\epsilon^{a^{\prime}b^{\prime}c^{\prime}}\,\int d^{4}xe^{ip\cdot x}
\langle 0|
\Bigg\{\mathrm{Tr}\Big[\gamma_{\mu} S_{u}^{dd^{\prime}}(x) 
\gamma_{\nu}S_c^{d^{\prime}d}(-x)\Big] 
\mathrm{Tr}\Big[\gamma_{\beta}\widetilde{S}_u^{aa^{\prime}}(x)\gamma_{\alpha}S_d^{bb^{\prime}}(x)]
\nonumber \\ 
&&-
\mathrm{Tr}\Big[\gamma_{\mu} S_{u}^{dd^{\prime}}(x) 
\gamma_{\nu}S_c^{d^{\prime}d}(-x)\Big]
\mathrm{Tr}\Big[\gamma_{\beta}\widetilde{S}_{u}^{ba^{\prime}}(x)\gamma_{\alpha}S_d^{ab^{\prime}}(x)\Big]
\Bigg\} \Big( \gamma^{\alpha}\gamma_{5}S_c^{cc^{\prime}}(x)\gamma_{5}\gamma^{\beta}\Big)\Bigg
|0 \rangle_\gamma.
 \label{eq:CorrF2Pc}
\end{eqnarray}
\end{widetext}
As a result, the QCD side of the correlation function in terms of the selected structures is obtained as
\begin{eqnarray}
\Pi^{QCD}_{\mu\nu}(p,q)&=&\Pi_{1}^{QCD}g_{\mu\nu}\pslash\eslash\qslash \,
+\Pi_{2}^{QCD}g_{\mu\nu}\eslash\qslash\,+
....
\end{eqnarray}

The following processes are applied as described in the previous subsection and magnetic dipole moment results are obtained in LCSR.
The  QCD and hadronic representations of the correlation function are then matched employing quark-duality assumption.   
By equating the coefficients of the structures $g_{\mu\nu}\pslash\eslash\qslash$ and $g_{\mu\nu}\eslash\qslash$, respectively for the $F_1$ and  $F_2$ we obtain sum rules for these two form factors. As a result, we get,
\begin{eqnarray}
\Pi^{Had}_{\mu\nu}(p,q)= \Pi^{QCD}_{\mu\nu}(p,q).
\end{eqnarray}
The explicit expressions of the LCSR for the $F_1^{Di}$ and  $F_2^{Di}$ are presented in the Appendix A.
We are now ready to move on to numerical analysis.

\section{Numerical analysis and discussions}\label{numerical}

This section is devoted to the numerical computations for the magnetic dipole moments of the $P_{c1}$, $P_{c2}$ and $P_{cs}$ pentaquark states. 
We use $m_u=m_d=0$, $m_s =96^{+8}_{-4}\,\mbox{MeV}$, $m_c = 1.275 \pm 0.02\,\mbox{GeV}$~\cite{Patrignani:2016xqp},  
$m_{P_{c1}} = 4440.3 \pm 1.3 ^{+4.1}_{-4.7} ~\mbox{MeV}$, 
$m_{P_{c2}}= 4457.3 \pm 0.6 ^{+4.1}_{-1.7} ~\mbox{MeV}$~\cite{Aaij:2019vzc}, 
$m_{P_{cs}}= 4458.8 \pm 2.7 ^{+4.7}_{-1.1}~\mbox{MeV}$~\cite{Aaij:2020gdg},
  $f_{3\gamma}=-0.0039~\mbox{GeV}^2$~\cite{Ball:2002ps}, 
$\langle \bar uu\rangle = 
\langle \bar dd\rangle=(-0.24 \pm 0.01)^3\,\mbox{GeV}^3$, $\langle \bar ss\rangle = 0.8\, \langle \bar uu\rangle$ $\,\mbox{GeV}^3$ \cite{Ioffe:2005ym},
$m_0^{2} = 0.8 \pm 0.1 \,\mbox{GeV}^2$ \cite{Ioffe:2005ym},  
$\langle g_s^2G^2\rangle = 0.88~ \mbox{GeV}^4$~\cite{Nielsen:2009uh},  
$\lambda_{P_{c1}}^{Di}=(1.44  \pm 0.23)\times 10^{-3}~\mbox{GeV}^6$, $\lambda_{P_{c2}}^{Di}=(3.02  \pm 0.48)\times 10^{-3}~\mbox{GeV}^6$~\cite{Wang:2019got},  $\lambda_{P_{cs}}^{Di}=(1.86 \pm0.31)\times 10^{-3}~\mbox{GeV}^6$ \cite{Wang:2020eep}, 
$\lambda_{P_{c1}}^{Mol}=(1.15^{+0.16}_{-0.18})\times 10^{-3}~\mbox{GeV}^6$, $\lambda_{P_{c2}}^{Mol}=(2.24^{+0.30}_{-0.34})\times 10^{-3}~\mbox{GeV}^6$ \cite{Chen:2020opr} and $\lambda_{P_{cs}}^{Mol}=1.13 \times 10^{-3}~\mbox{GeV}^6$. 
Another set of main input parameters are the photon wavefunctions of different twists, entering the DAs. These wavefunctions are given in Ref. \cite{Ball:2002ps}.

\begin{widetext}

  \begin{figure}[htp]
\centering
  \includegraphics[width=0.45\textwidth]{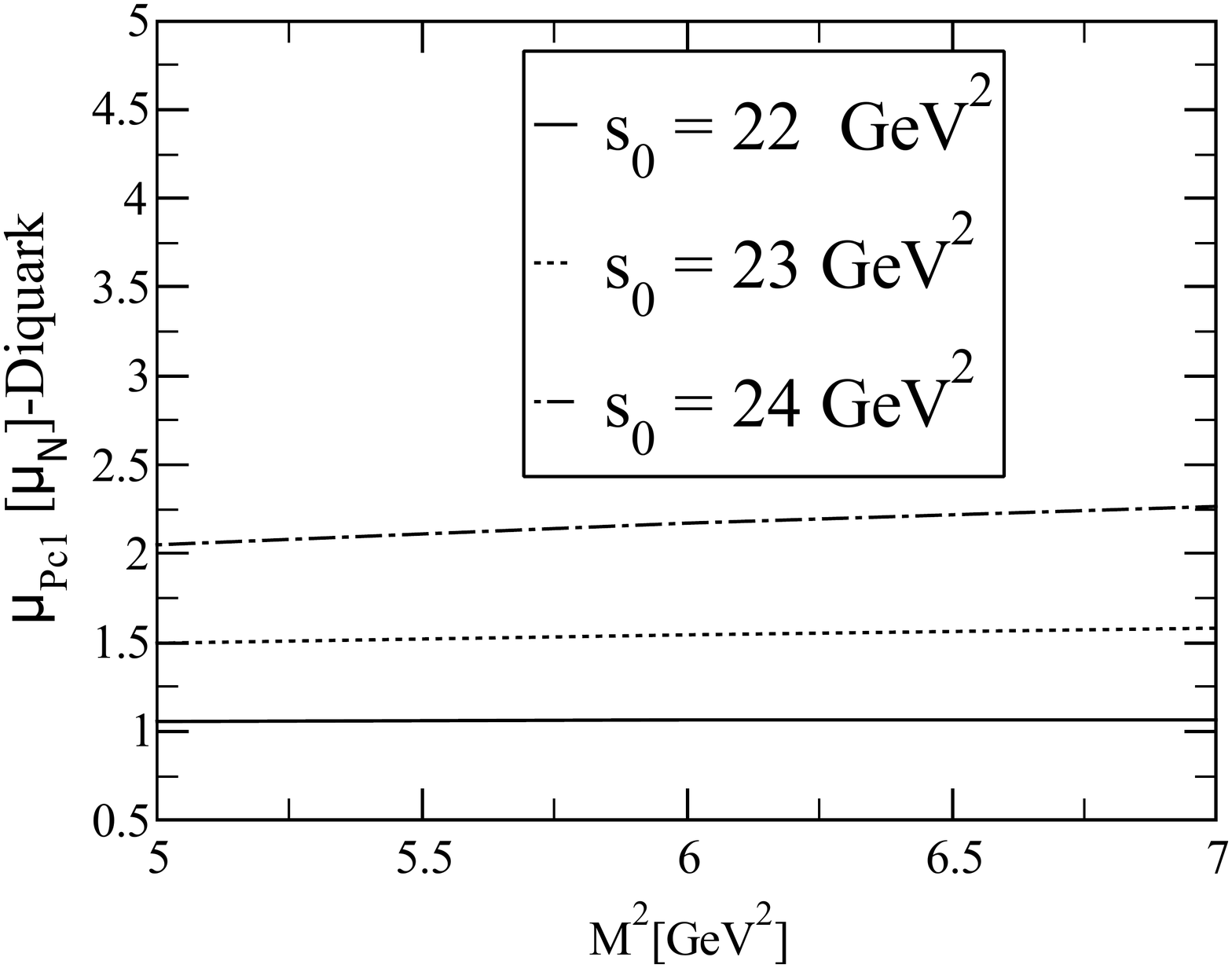}~~~
  \includegraphics[width=0.45\textwidth]{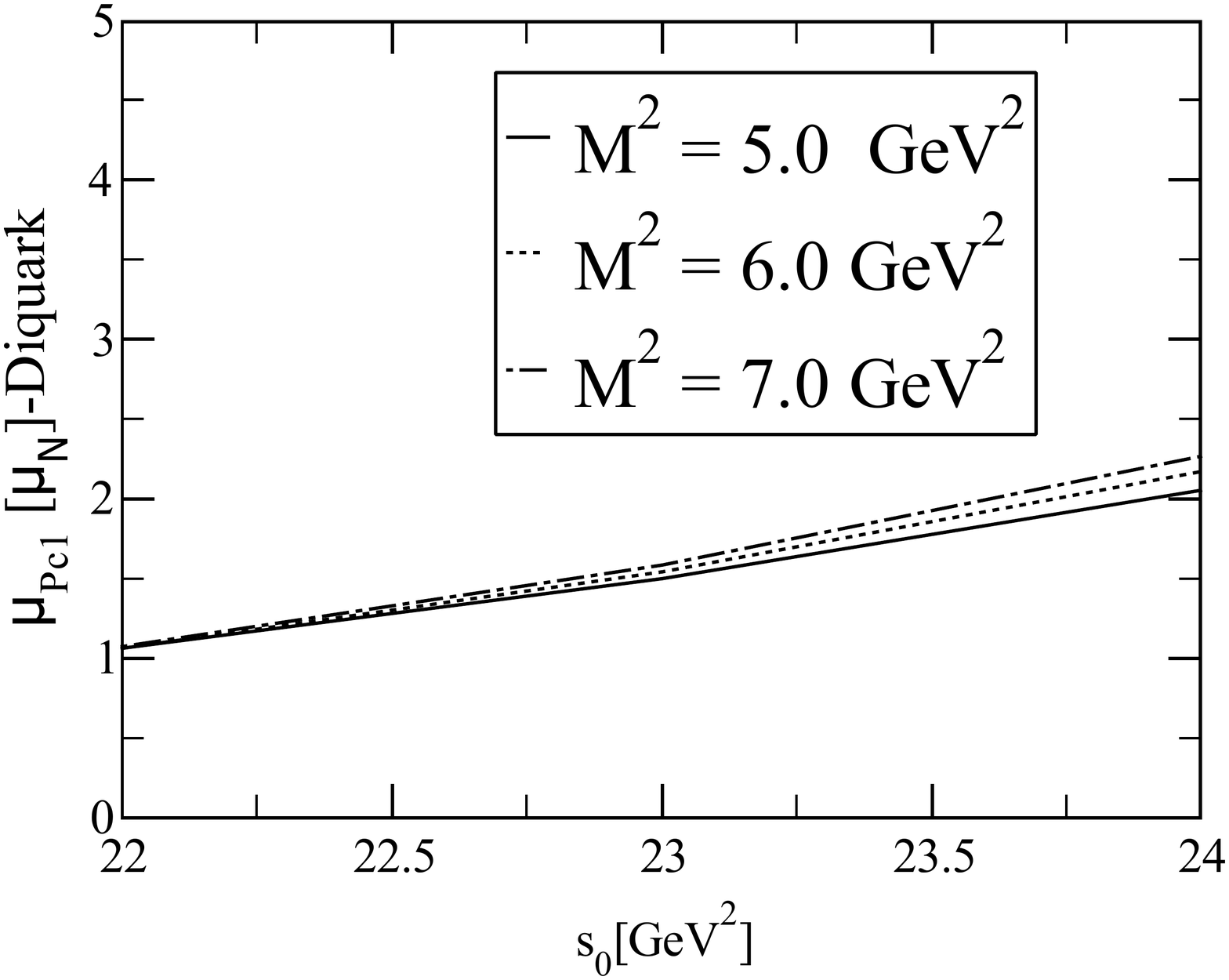}\\
  \vspace{0.5cm}
  \includegraphics[width=0.45\textwidth]{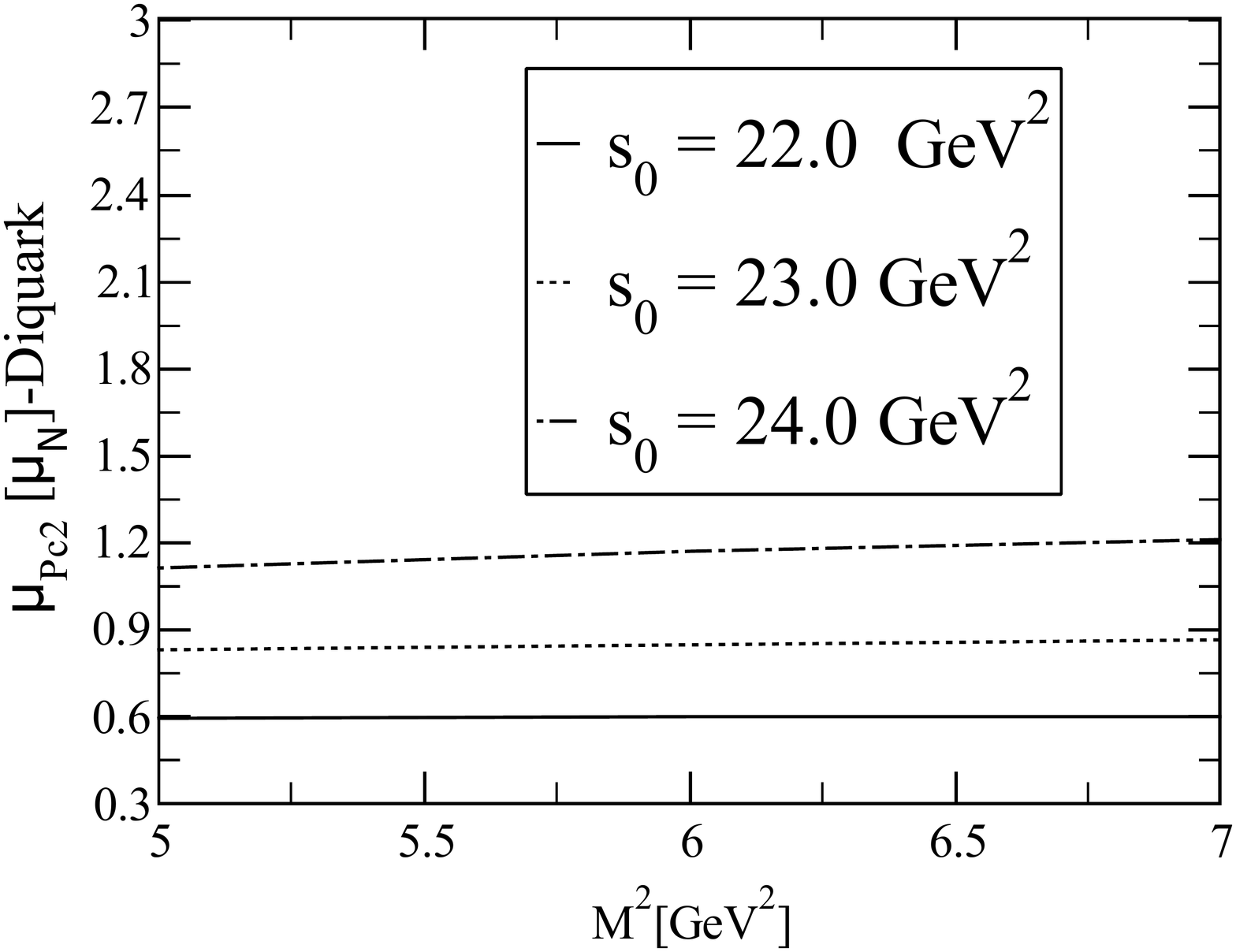}~~~
  \includegraphics[width=0.45\textwidth]{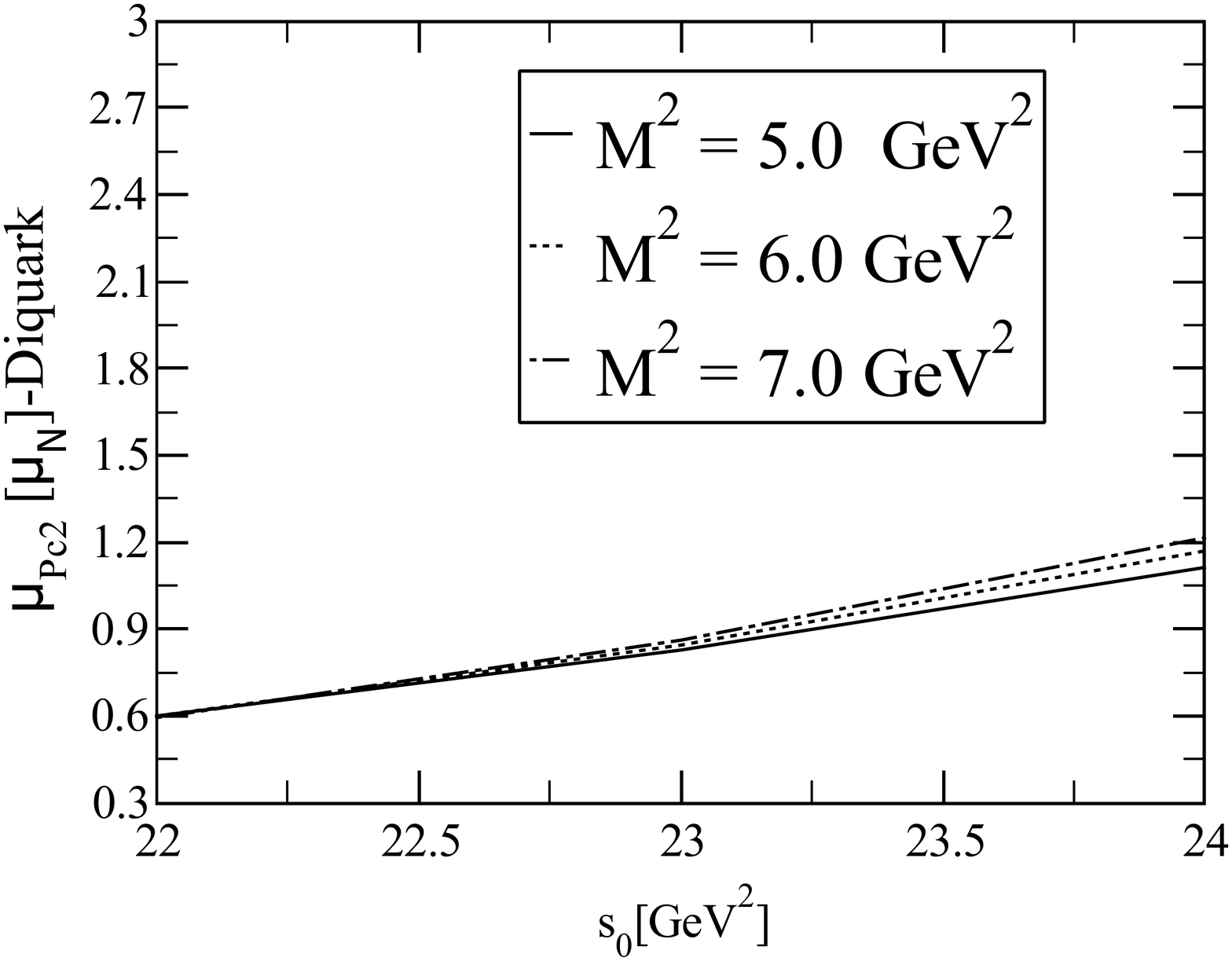}\\
  \vspace{0.5cm}
  \includegraphics[width=0.45\textwidth]{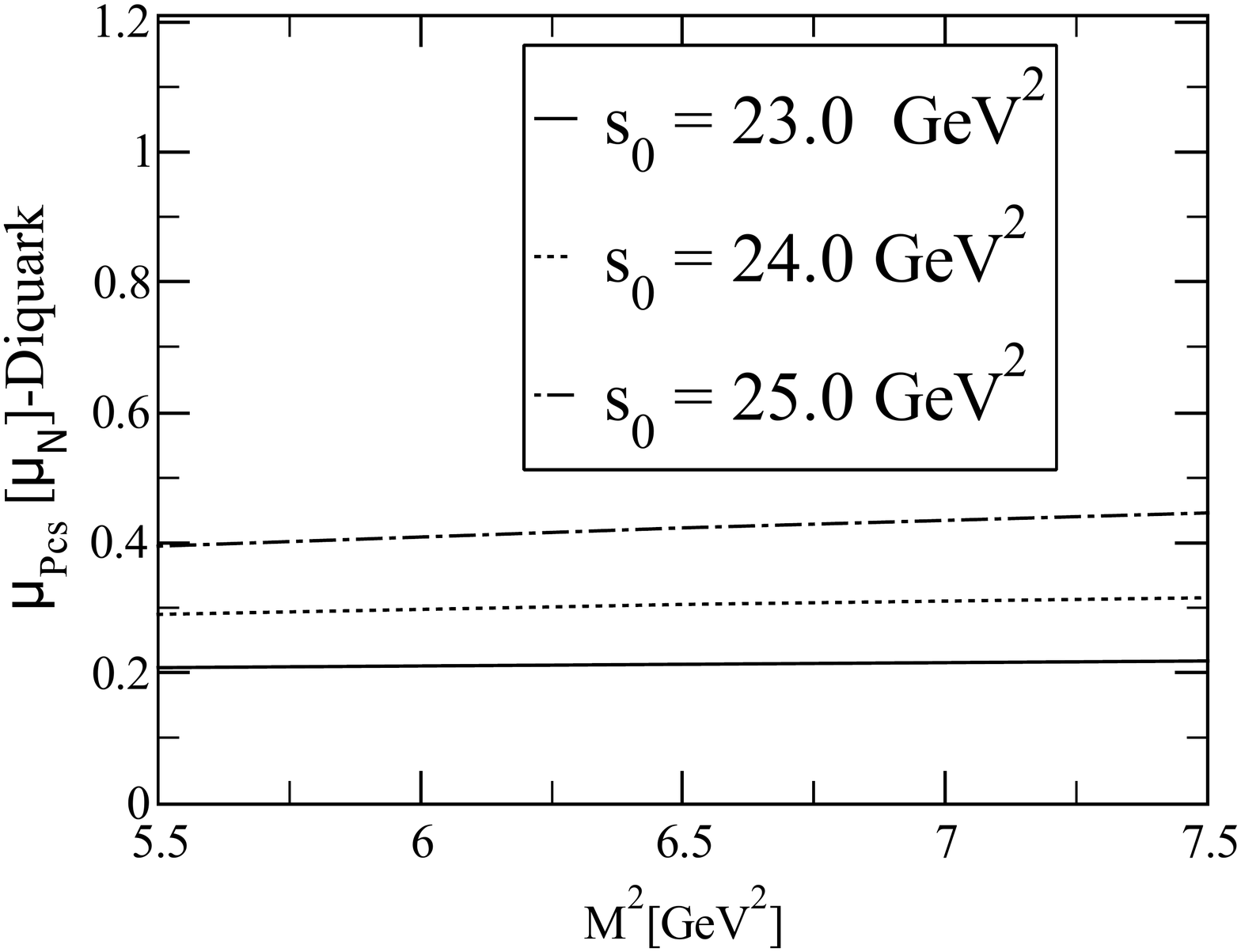}~~~
  \includegraphics[width=0.45\textwidth]{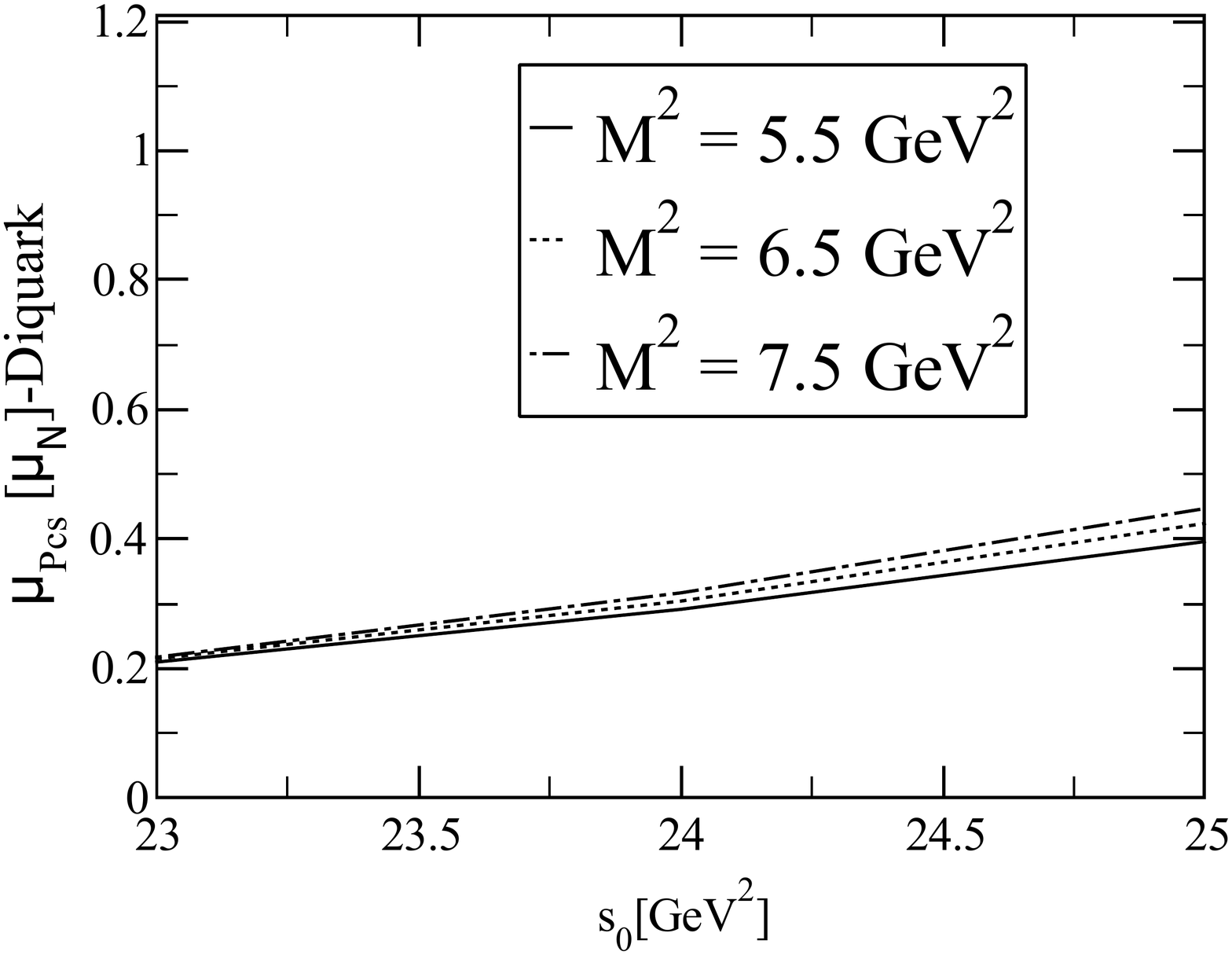}
 \caption{Variations of the magnetic dipole moments of  $\mu_{P_{c1}}$,  $\mu_{P_{c2}}$ and $\mu_{P_{cs}}$in diquark-diquark-antiquark picture with $M^2$ and $s_0$.}
    \label{s0Mfig1}
  \end{figure}
 
  \end{widetext}

Except the above mentioned input parameters, the estimations for the magnetic dipole moments of pentaquark states depend on two auxiliary parameters: Borel mass parameter $ M^2 $ and continuum threshold $ s_0 $. 
 %
 %
 According to the philosophy of the method used, the observables  under examination should be weakly dependent on the variations of these auxiliary parameters. 
 The continuum threshold is considered to be the point where the excited states and continuum begin to contribute to the correlation function. 
 The upper and lower bound of the Borel parameter is decided by demanding that both the contributions of the higher states and continuum are adequately suppressed and the contributions coming from higher dimensional terms are small.
Our numerical analysis leads to the conclusion that these requirements are fulfilled in the regions shown below for the considered pentaquark states.
\begin{align*}
&22.0~\mbox{GeV}^2 \leq s_0 \leq 24.0~\mbox{GeV}^2~ P_{c1}~ \mbox{and} ~ P_{c2}~\mbox{states},\\
&23.0~\mbox{GeV}^2 \leq s_0 \leq 25.0~\mbox{GeV}^2 ~  \mbox{for}~ P_{cs}~ \mbox{state},\\
\nonumber\\
&5.0~\mbox{GeV}^2 \leq M^2 \leq 7.0~\mbox{GeV}^2 ~ \mbox{for} ~ P_{c1}~ \mbox{and} ~ P_{c2}~\mbox{states},\\
&5.5~\mbox{GeV}^2 \leq M^2 \leq 7.5~\mbox{GeV}^2 ~ \mbox{for} ~ P_{cs}~ \mbox{state}.
\end{align*}
By having the values of all input parameters, we can start carrying out numerical computations.
In Fig.~\ref{s0Mfig1}, as an example, we depict the dependencies of the magnetic dipole moments in diquark-diquark-antiquark picture on $M^2$ and $s_0$. As is seen, the deviation of the results in connection with the $s_0$ is remarkable however there is much less dependence of the physical observables under consideration on the $M^2$ in its working window.

%
Our final results for the magnetic dipole moments are presented in Table. \ref{table}. 
\begin{table}[htp]
	\addtolength{\tabcolsep}{10pt}
	\begin{center}
\begin{tabular}{lccccc}
	   \hline\hline
	   Picture&  $\mu_{P_{c1}}$	& $\mu_{P_{c2}}$		&$\mu_{P_{cs}}$   \\
	   \hline\hline
	   Diquark&   $ 1.62^{+0.65}_{-0.57}$     &$ 0.88^{+0.32}_{-0.29}$ 	&$ 0.34^{+0.13}_{-0.11}$\\
	   Molecule&  $ 3.49^{+1.49}_{-1.31}$     &$2.78^{+0.94}_{-0.83}$ 	&$ 1.75^{+0.64}_{-0.58}$\\
	   \hline\hline
\end{tabular}
\end{center}
\caption{Results of the magnetic dipole moments (in units of $\mu_N$) for the pentaquark states.}
	\label{table}
\end{table}
%
Our results cover errors originated from the uncertainty of the determinations of auxiliary parameters ($M^2$ and $s_0$) and other input parameters used in the analyses.
The numerical values of the magnetic dipole moments acquired by means of two configurations differ considerably from each other, which can be used to determine the fundamental structure of these  pentaquark states.
That is to say, as many theoretical models give compatible results on the decay channels and spectroscopic parameters with the experimental data preventing us assigning precise substructure for pentaquarks, the experimental measurement of the magnetic dipole moments of these pentaquarks indeed can help us exactly distinguish their substructures. 
Since there are no theoretical result and experimental data for the results we obtained for these pentaquark states, we cannot compare our results.
However, we may compare the result of the $P_{cs}$ state with the  $P_c (4312)$ state magnetic dipole moments.
It can be seen that the quark configurations of the $P_{cs}$ and $P_c (4312)$ particles are similar and therefore the magnetic dipole moment results of these particles can be expected to be close to each other.
In Ref. \cite{Ozdem:2021btf}, the magnetic dipole moment of $P_c(4312)$  in the diquark-diquark-antiquark and molecular pictures are extracted as $\mu_{P_c}= 0.40 \pm 0.15\, \mu_N$ and  $\mu_{P_c}= 1.98 \pm 0.75\, \mu_N$, respectively.
 The magnetic dipole moment of the $P_c(4312)$ pantaquark state has also been extracted via the QCD sum rule and its extension in the weak
electromagnetic field by employing a molecular type interpolating current in Ref. \cite{Xu:2020flp}. The numerical value was obtained as $\mu_{P_c}= 0.59^{+10}_{-0.20}$. 
As one can see from these estimations, the numerical values for the magnetic dipole moments of $P_{c}(4312)$ and $P_{cs}$ states acquired in the present study are close to each other and we see a reasonable SU(3) flavor violation which is roughly $\% 15$.
The obtained results in both pictures are considerably different compared to the result of Ref. \cite{Xu:2020flp}.
Comparing the results obtained using different theoretical models with our results can give an idea about the consistency of our predictions.

 As a by product, we also obtain the electric quadrupole ($Q_{P_{c1}}$) and magnetic octupole ($O_{P_{c1}}$) moments of the $P_{c1}$ pentaquark as
 \begin{align}
  &Q_{P_{c1}}^{Di}= (2.0^{+0.6}_{-0.5})\times 10^{-2}~ \mbox{fm$^2$}, ~~~~~~~~
  Q_{P_{c1}}^{Mol}= 0.28^{+0.09}_{-0.07}~ \mbox{fm$^2$}, \\
  \nonumber\\
  &O_{P_{c1}}^{Di}= (0.24^{+0.07}_{-0.06})\times 10^{-3} ~ \mbox{fm$^3$}, ~~~~~~
  O_{P_{c1}}^{Mol}= (4.83^{+1.36}_{-1.13})\times 10^{-3} ~ \mbox{fm$^3$}.
 \end{align}
Just like the magnetic dipole moment results, we can see that the electric quadrupole and magnetic octupole moments results obtained by using two different pictures are different from each other.  The $Q_{P_{c1}}$ and $O_{P_{c1}}$ of the $P_{c1}$ pentaquark state showing a non-spherical charge distribution.
 

 In summary, the observation of new pentaquark states such as $P_c(4312)$, $P_c(4440)$,  $P_c(4457)$ and $P_{cs}(4459)$, ensures a new platform to investigate the exotic states in QCD. There are different interpretations of their inner structures and quantum numbers, and these should be shedded light on with further research.
In the present work, stimulated by the observation of the hidden-charm pentaquark states we have achieved the magnetic dipole moments of the  $P_c(4440)$,  $P_c(4457)$  and  $P_{cs}(4459)$ by considering them as diquark-diquark-antiquark and molecular pictures with quantum numbers $J^{P} =\frac{3}{2}^{-}$,  $J^{P} =\frac{1}{2}^{-}$ and $J^{P} =\frac{1}{2}^{-}$ by means of the light-cone QCD sum rule, respectively.
As a by product, the electric quadrupole and magnetic octupole moments of the $P_c(4440)$ pentaquark have also extracted. 
Our predictions for electromagnetic multi-pole moments of pentaquark states  may be checked via different theoretical models such as Lattice QCD, chiral perturbation theory etc.
With the latest developments in the experimental side, we hope that we will be able to measure the magnetic dipole moment of newly formed multi-quark states, especially pantaquark states in the future.
Any experimental measurements of the electromagnetic multi-pole moments of the hidden-charm pentaquark states and comparison of the obtained results with the predictions of the this study may provide as helpful knowledge on the internal structure of the these states as well as the non-perturbative behaviors of the strong interaction at the low-energy region.

\section{Acknowledgments}

We thank Hua-Xing Chen for providing us the numeric value of the residue of the $P{cs}(4459) $ state.

 \begin{widetext}
\section*{Appendix A: Explicit forms of the  \texorpdfstring{$F_1^{Di}$}{} and \texorpdfstring{$F_2^{Di}$}{}  functions } 
In this appendix, we give the explicit expressions for the  $F_1^{Di}$ and $F_2^{Di}$ functions: 
\begin{align}
 F_1^{Di}&= \frac{e^{\frac{m^2_{P_{c1}}}{M^2}}}{\lambda^{2-Di}_{P_{c1}}}\Bigg\{
 \frac{\langle g_s^2 G^2 \rangle}{108716359680\,\pi^7}\Bigg[20\,\pi^2\,f_{3\gamma}
 \Big(4\,(32e_d+41e_u)\,I[0,4,4,0]+3\,(9e_d+20e_u)\,I[0,4,5,0]\Big)I_2[\mathcal{V}]\nonumber\\
 &+(72e_d+720e_u-702e_c)\,I[0,5,2,1]-(324e_d+2184e_u-2673e_c)\,I[0,5,2,2]+(492e_d+2456e_u-3511e_c)\nonumber\\
 &\times I[0,5,2,4]-(300e_d+1240e_u-1811e_c)\,I[0,5,2,4]+(60e_d+248e_u-271e_c)\,I[0,5,2,5]+(216e_d+2160e_u\nonumber\\
 & -2106)\,I[0,5,3,1]+(684e_d+4728e_d-5693e_c)\,I[0,5,3,2]-(624e_d+3424e_u-4483e_c)\,I[0,5,3,3]+(156e_d\nonumber\\
 &+856e_u-893e_c)\,I[0,5,3,4]+(216e_d+2160e_u-2106e_c)\,I[0,5,4,1]-(396e_d+2904e_u-3375e_c)\nonumber\\
 &\times I[0,5,4,2] +(132e_d+968e_u-973e_c)\,I[0,5,4,3]
 -(72e_d+720e_u-702e_c)\,I[0,5,5,1]+(36e_d+360e_u\nonumber\\
 &-351e_c)\,I[0,5,5,2]\Bigg]\nonumber\\
 &-
 \frac{m_c\,\langle \bar qq \rangle}{3774873\,\pi^5}\Bigg[192\,e_c\,
 \Big(I[0,5,2,2]-2\,I[0,5,2,3]+I[0,5,2,4]-2\,I[0,5,3,2]+2\,I[0,5,3,3]+I[0,5,4,2]\Big)\nonumber\\
 &-40\,\pi^2\,f_{3\gamma}(e_d+10e_u)\,
 I_2[\mathcal{V}]\,I[0,4,4,0]+3(e_d+14e_u)\,I_4[\tilde S]\, I[0,5,4,0]
 \Bigg]\nonumber\\
 &+\frac{f_{3\gamma}}{3019898880\,\pi^5}(13e_d+58e_u)\,I_2[\mathcal{V}]\,I[0,6,5,0]\nonumber\\ 
 &-\frac{e_c}{880803840\,\pi^7}\Bigg[4\,I[0,7,2,3]-13\,I[0,7,2,4]+15\,I[0,7,2,5]
 -7\,I[0,7,2,6]+I[0,7,2,7]-12\,I[0,7,3,3]\nonumber\\
 &+27\,I[0,7,3,4]-18\,I[0,7,3,5]+3\,I[0,7,3,6]+12\,I[0,7,4,3]+3\,I[0,7,4,5]-4\,I[0,7,5,3]+I[0,7,5,4]\Bigg]\Bigg\},
 \\
 F_2^{Di}&= \frac{m_{P_{c1}} \,e^{\frac{m^2_{P_{c1}}}{M^2}}}{\lambda^{2-Di}_{P_{c1}}}\Bigg\{ \frac{\langle g_s^2 G^2 \rangle}{108716359680\,\pi^7}\Bigg[
    20\, f_{3\gamma} \pi^2\, \Big\{-4 \Big((52 e_d - 19 e_u) I_2[\mathcal A] - 
      9 (32 e_d + 41 e_u) I_2[\mathcal V] + 4 (3 e_d + 2 e_u)
      \nonumber\\
      & \times I_6[\psi^\nu]\Big)\,I[0, 4, 4, 0]
      +3 \Big(8 (-34 e_d + e_u) I_2[\mathcal A] + 9 (9 e_d + 220 e_u)\, I_2[\mathcal V]\Big)\,I[0, 4, 5, 0]\Big\}
      -\Big\{
      ( 702 e_c-72 e_d - 720 e_u) \nonumber\\
      & \times I[0, 5, 2, 1]+ (324 e_d - 2673 e_c + 2184 e_u)\,I[0, 5, 2,2] 
    + (-492 e_d + 3511 e_c - 2456 e_u)\,I[0, 5, 2, 
   3] + (300 e_d  \nonumber \\
    & - 1811 e_c + 1240 e_u)\,I[0, 5, 2,4] + (-60 e_d + 271 e_c - 248 e_u)\,I[0, 5, 2, 
   5] + (216 e_d - 2106 e_c + 2160 e_u)\,I[0, 5, 3, 
   1] \nonumber\\
   & + (-684 e_d + 5697 e_c - 4728 eu) I[0, 5, 3,2] + (624 e_d - 4484 e_c + 3424 e_u)\,I[0, 5, 3, 3]  + (893 e_c-156 e_d - 856 e_u) \nonumber\\
   & 
   \times I[0, 5, 3, 4] + (-216 e_d + 2106 e_c - 2160 e_u)\,I[0, 5, 4, 1] + (396 e_d - 3375 e_c + 2904 e_u)\,I[0, 5, 4, 
   2]   + (-132 e_d \nonumber\\
    &  + 973 e_c - 968 e_u)
    I[0, 5, 4, 3] + (72 e_d - 702 e_c + 720 e_u)\,I[0, 5, 5, 
   1] + (-36 e_d 
   + 351 e_c - 360 e_u)\,I[0, 5, 5, 2]
      \Big\}\Bigg]\nonumber\\
        & -\frac{m_c\,\langle \bar qq \rangle}{125829120\,\pi^5}\Bigg[
 -120\, (e_d + 10 e_u)\, f_{3\gamma}\, \pi^2\, I_2[\mathcal V]\, I[0, 4, 4, 0]
+ (3 e_d + 14 e_u) I_4[\mathcal S]+ 2 e_d\, I_4[\mathcal T_1] I[0, 5, 4, 0]\nonumber\\
 &+64\,e_c\,\Big(
  I[0, 5, 2, 2] - 2\,I[0, 5, 2, 3] + I[0, 5, 2, 4] - 
    2\,I[0, 5, 3, 2] + 2\,I[0, 5, 3, 3] + I[0, 5, 4, 2] \Big)\Bigg]\nonumber\\
   &+\frac{f_{3\gamma}}{3019898880\,\pi^5}    (13 e_d + 58 e_u) I_2[\mathcal V]\,I[0, 6, 5, 0]\nonumber
   \\
   &-\frac{e_c}{880803840\, \pi^7}\Bigg[
   4\,I[0, 7, 2, 3] - 13\,I[0, 7, 2, 4] + 15\,I[0, 7, 2, 5] - 
 7\,I[0, 7, 2, 6] + I[0, 7, 2, 7] - 12\,I[0, 7, 3, 3]\nonumber\\
 &+ 
 27\,I[0, 7, 3, 4] - 18\,I[0, 7, 3, 5] + 3\,I[0, 7, 3, 6] + 
 12\,I[0, 7, 4, 3] - 15\,I[0, 7, 4, 4] + 3\,I[0, 7, 4, 5] - 
 4\,I[0, 7, 5, 3]\nonumber\\
 &+ I[0, 7, 5, 4]\Bigg]\Bigg\}.
\end{align}
We should also point out that in the above expressions, for simplicity we have only given the terms that give important contributions to the numerical values of the magnetic moments, and have not presented many higher dimensional contributions, although they have been considered in the numerical calculations.

The functions~$I[n,m,l,k]$, $I_1[\mathcal{A}]$,~$I_2[\mathcal{A}]$,~$I_3[\mathcal{A}]$,~$I_4[\mathcal{A}]$,
~$I_5[\mathcal{A}]$, and ~$I_6[\mathcal{A}]$ are
defined as:
\begin{align}
 I[n,m,l,k]&= \int_{4 m_c^2}^{s_0} ds \int_{0}^1 dt \int_{0}^1 dw~ e^{-s/M^2}~
 s^n\,(s-4\,m_c^2)^m\,t^l\,w^k,\nonumber\\
 I_1[\mathcal{A}]&=\int D_{\alpha_i} \int_0^1 dv~ \mathcal{A}(\alpha_{\bar q},\alpha_q,\alpha_g)
 \delta'(\alpha_ q +\bar v \alpha_g-u_0),\nonumber\\
  I_2[\mathcal{A}]&=\int D_{\alpha_i} \int_0^1 dv~ \mathcal{A}(\alpha_{\bar q},\alpha_q,\alpha_g)
 \delta'(\alpha_{\bar q}+ v \alpha_g-u_0),\nonumber\\
   I_3[\mathcal{A}]&=\int D_{\alpha_i} \int_0^1 dv~ \mathcal{A}(\alpha_{\bar q},\alpha_q,\alpha_g)
 \delta(\alpha_ q +\bar v \alpha_g-u_0),\nonumber\\
   I_4[\mathcal{A}]&=\int D_{\alpha_i} \int_0^1 dv~ \mathcal{A}(\alpha_{\bar q},\alpha_q,\alpha_g)
 \delta(\alpha_{\bar q}+ v \alpha_g-u_0),\nonumber\\
   I_5[\mathcal{A}]&=\int_0^1 du~ A(u)\delta'(u-u_0),\nonumber\\
 I_6[\mathcal{A}]&=\int_0^1 du~ A(u),\nonumber
 \end{align}
 where $\mathcal{A}$ represents the corresponding photon distribution amplitudes.
\end{widetext}

\bibliography{PentaquarksMMr}
\end{document}